\documentclass[traditabstract]{aa} 
\usepackage{graphicx}
\usepackage{txfonts}
\usepackage{natbib}
\usepackage{lscape}
\newcommand{\reduceme}{\mbox{R\raisebox{-0.35ex}{E}D%
\hspace{-0.05em}\raisebox{0.85ex}{uc}\hspace{-0.90em}%
\raisebox{-.35ex}{{m}}\hspace{0.05em}E}}

\begin{document}
   \title{Formation and evolution of dwarf early-type galaxies in the Virgo cluster}

   \subtitle{II. Kinematic Scaling Relations}

   \author{E. Toloba \inst{1,2,3}\thanks{Fulbright Fellow} \and
    A. Boselli \inst{4} \and
  R. F. Peletier \inst{5} \and
 J. Falc\'{o}n-Barroso \inst{6,7} \and
G. van de Ven \inst{8} \and
   J. Gorgas \inst{1}}
  \institute{Departamento de Astrof\'{i}sica y CC. de la Atm\'{o}sfera, Universidad Complutense de Madrid,
              28040, Madrid, Spain\\
              \email{toloba@ucolick.org} 
              \email{jgorgas@fis.ucm.es}
 \and UCO/Lick Observatory, University of California, Santa Cruz, 1156 High Street, Santa Cruz, CA 95064
  \and Observatories of the Carnegie Institution of Washington, 813 Santa Barbara Street, Pasadena, CA 91101
         \and Laboratoire d'Astrophysique de Marseille-LAM, Universit\'e d'Aix-Marseille \& CNRS, UMR 7326, 38 rue F. Joliot-Curie, 13388 Marseille Cedex 13, France\\
               \email{alessandro.boselli@oamp.fr} 
\and Kapteyn Astronomical Institute, Postbus 800, 9700 AV Groningen, the Netherlands\\
\email{peletier@astro.rug.nl} 
\and Instituto de Astrof\'{i}sica de Canarias, V\'{i}a
 L\'{a}ctea s$/$n, La Laguna, Tenerife, Spain.
\and Departamento de Astrof\'{i}sica, Universidad de La Laguna, 
 E-38205, La Laguna, Tenerife, Spain.\\
\email{jfalcon@iac.es}
\and Max Planck Institute for Astronomy, K$\ddot{\rm o}$nigstuhl 17, 69117 Heidelberg, Germany\\
\email{glenn@mpia.de}
             }

   \date{Received ; accepted }

\abstract{ We place our sample of 18 Virgo dwarf early-type galaxies (dEs) on the $(V-K)_{e}$-velocity dispersion, Faber-Jackson, and Fundamental Plane (FP) scaling relations for massive early-type galaxies (Es). We use a generalized velocity dispersion, which includes rotation, to be able to compare the location of both rotationally and pressure supported dEs with those of early and late-type galaxies. We find that dEs seem to bend the Faber-Jackson relation of Es to lower velocity dispersions, being the link between Es and dwarf spheroidal galaxies (dSphs). Regarding the FP relation, we find that dEs are significantly offset with respect to massive hot stellar systems, and re-casting the FP into the so-called $\kappa$-space suggests that this offset is related to dEs having a total  mass-to-light ratio higher than Es but still significantly lower than dSph galaxies. Given a stellar mass-to-light ratio based on the measured line indices of dEs, the FP offset allows us to infer that the dark matter fraction within the half light radii of dEs is on average $\gtrsim 42\%$ (uncertainties of 17$\%$ in the $K$ band and 20$\%$ in the $V$ band), fully consistent with an independent estimate in an earlier paper in this series. We also find that dEs in the size-luminosity relation in the near-infrared, like in the optical, are offset from early-type galaxies, but seem to be consistent with late-type galaxies. We thus conclude that the scaling relations show that dEs are different from Es, and that they further strengthen our previous findings that dEs are closer to and likely formed from late-type galaxies. }
           
\keywords{Galaxies: clusters: individual: Virgo
                Galaxies: dwarf
                Galaxies: elliptical and lenticular, cD
                Galaxies: kinematics and dynamics
                Galaxies: formation
                Galaxies: dark matter
               }

   \maketitle
%

\section{Introduction}

Around $75\%$ of the total baryonic matter in the local universe is in form of stars in spheroidal objects \citep[early-type galaxies and bulges of spiral galaxies,][]{Fukugita98,Renz06}. Within them, dwarf early-type galaxies (dEs, objects with M$_B \geq -18$ mag including ellipticals and lenticulars) are by far the most numerous systems \citep{FB94}, but, their origin is not well known. Whether they are the low-luminosity extension of massive early-type galaxies (Es, objects with M$_B \leq -18$ mag) or they are a different population of galaxies is still on active debate. Observational evidence seems to suggest that these faint systems form a heterogeneous population, more similar to late-type galaxies than to ellipticals. They present a great variety of underlying structures, like discs, spiral arms, irregular features, some of them are also nucleated \citep{Lisk06b,Lisk06a}. Their luminosity profiles are similar to exponential \citep{Caon93,Ferrarese06,Lisk07} and they are not composed of a simple, old and metal rich stellar population, but they rather span luminosity-weighted ages from 1 Gyr to as old as the oldest objects in the universe \citep[][]{Mich08,Koleva09}. Although in appearance dEs look like Es, their physical properties seem to be different from those of massive early-types.

It has been proposed that the mechanisms that formed these systems are purely internal, i.e. the kinetic energy generated in supernovae sweeps away the gas of the galaxy leaving it without any fuel to create new stars \citep{YosAri87}. These processes are not expected to depend on the properties of the surrounding environment. They can not explain the morphology segregation observed also for dwarf galaxies; dEs are the most common population in dense environments, like clusters, whereas late-type low-luminosity galaxies dominate in number in the field \citep{Dress80,Sand85,FB94,Blant05,Crot05}. This suggests that the environment plays a key role in shaping the evolution of these low-luminosity stellar systems, possibly transforming late-type galaxies into dEs \citep[e.g.][]{Boselli08a}. Therefore external processes related to the perturbations induced on galaxies by the environment where they reside, could be responsible of the formation of dEs. These mechanisms can be purely gravitational, galaxy harassment,  \citep[][]{Moore98,Mast05}, or due to the interaction with the intergalactic medium, ram pressure stripping,  \citep[][]{Boselli08a}. To gravitationally disrupt late-type low-luminosity galaxies and transform them into dEs, close encounters are required. However, in clusters like Virgo, the density of objects needed to make this process efficient is only achieved in the center, otherwise the time necessary for multiple encounters to occur is too long \citep{BG06}. On the contrary, in a ram pressure stripping event, late-type galaxies infalling into the cluster lose their gas and become quiescent galaxies just through their interaction with the intergalactic medium (IGM) \citep{Boselli08a,Boselli08b}. 

The most direct way to discriminate between these two different scenarios, harassment versus ram pressure, is to analyse the kinematic properties of dwarf early-type galaxies. 
In a ram pressure scenario the angular momentum is conserved long after the perturbation, thus the ratio between the mean velocity and the velocity dispersion ($v_{max}/\sigma$) remains roughly constant. In case of harassment, after several nearby encounters, the systems are rapidly heated increasing so that $v_{max}/\sigma$ decreases.
In the last decade a significant effort has been made to study the kinematic properties of dEs. Several works have shown that some dEs show significant rotation while others do not \citep{Ped02,Geha02,Geha03,VZ04,Chil09,etj09b,etj10}. The lack of a statistically significant sample, however, prevented these authors to pose strong constraints on the evolution of this class of objects.

In order to understand the origin of these dwarf systems, we have embarked on an ambitious programme to obtain radial kinematic information for a large sample of dEs in the Virgo cluster. 
In \citet[][hereafter T09]{etj09b} we have shown that the kinematic properties of dEs are directly related to the Virgocentric distance: rotationally supported dEs tend to be found in the outer parts of the cluster while pressure supported systems are preferentially located in the center. Moreover, those dEs in the outer parts are younger than those in the center and present disky underlying structures which are lacking in those located in the cluster core. In \citet[][hereafter Paper I]{etj10} we have shown that rotationally supported dEs have rotation curves similar to those of late-type galaxies of comparable luminosity. All these evidences are consistent with a scenario where low-luminosity late-type galaxies, the major population in the field, enter into the Virgo cluster where they are transformed into dEs. Those dwarf early-types currently in the outer parts of the Virgo cluster are likely to be ram-pressure stripped late-type galaxies, whereas those in the center, could have suffered more than one transforming mechanism. A full discussion on this topic is presented in T09 and Paper I. 

For a complete and comprehensive study of this population of galaxies we analyse here their kinematic scaling relations: the Faber-Jackson, the Fundamental Plane and the colour-velocity dispersion relations. These diagrams,  widely used to determine distances of early-type galaxies due to their small scatter, are a useful tool to study and compare at the same time the properties of low and high-luminosity galaxies. 
The tightness of these scaling relations has been classically interpreted as an indication that early-type galaxies constitute a homogeneous population shaped by the same formation processes \citep[e.g.][]{Jorgensen96,P98}. The location of dwarf galaxies in these relations might thus give information about their origin. 

Some works have addressed the scaling relations of dwarf galaxies in the past \citep[e.g.][]{Bender92,Graham03,Geha03,DR05,Matkovic05,Kormendy09}, but whether dEs follow the same trends as Es is still unclear. 
While \citet{Kormendy09} claim that Es and dEs are two different populations of galaxies finding that they follow perpendicular trends in, e.g., the Kormendy relation, other works such as \citet{Graham03} or \citet{Ferrarese06} find that there is a continuity in the physical properties of dwarf and massive early-types. The debate is still open.

Kinematic scaling relations can also give information on the dark matter content in galaxies by constraining the dynamical mass-to-light ratio ($M_{\rm dyn}/L$) within the half light radius \citep{Zaritsky06,Zaritsky08,Zaritsky11}. More common approaches to determine the $M_{\rm dyn}/L$, either through a virial mass estimate or dynamical modeling, have shown that Es within their half-light radii are dominated by baryons \citep[e.g.][]{Cappellari06} whereas the large  $M_{\rm dyn}/L$ values for dwarf spheroidal galaxies (dSphs) show that dark matter dominates in these systems \citep[e.g.][]{Wolf10}. Adopting also virial mass estimates, we showed in Paper I that dEs have larger $M_{\rm dyn}/L$ than Es but smaller than dSphs, suggesting that some dark matter is present in these systems. Here, we use the Fundamental Plane to obtain an independent constraint on the dark matter fraction within the half-light radii of our dEs.

All these studies have been done mainly in the optical bands, which do not trace the bulk of the stellar populations and are biased towards the young stars present in these galaxies. Here, we go to the near-infrared, a wavelength range that traces the stellar mass in galaxies \citep{Gavazzi96,BelldeJong01}, is insensitive to recent episodes of star formation and is much less affected by extinction. 
Another novelty of this work is the way the velocity dispersion is computed for dwarf galaxies. As previously suggested \citep{Zaritsky06,Zaritsky08,FB11}, the velocity dispersion is best measured within the half light radius including both the rotation and the random motions of the stars, so that it represents the kinetic energy of the systems.

This paper is structured as follows: Section \ref{FP_sample} presents the sample and briefly describe the observations and the reduction procedure. In Section \ref{FP_phot} we define the photometric parameters measured for this study and compare them with the literature.
In Section \ref{scaling_rels} we analyse the scaling relations of early-type galaxies. In Section \ref{FP_disc} we discuss and summarise the conclusions about the possible origin of dEs. Appendix \ref{app_errors} develops the methodology followed to measure the photometric errors. Appendix \ref{app_phot_rels} shows, for completeness, the Kormendy relation, the colour-magnitude and the colour-size diagrams.

Throughout this paper we use a Hubble constant of $H_0=73$ km s$^{-1}$ Mpc$^{-1}$ \citep{Mei07}, which corresponds to a distance for Virgo of 16.7 Mpc.


\section{Data}\label{FP_sample}

\subsection{The sample}

\begin{table*}
\begin{center}
\caption{Sample of dEs in the Virgo cluster. \label{sample}}
{\renewcommand{\arraystretch}{1.3}
\begin{tabular}{|c|c|c|c|c|c|c|}
\hline \hline
Galaxy              &  RA(J2000) & Dec.(J2000)  & Type & $(v_{max}/\sigma)^*$ & $\sigma_e$      & Age          \\
                        &  (h:m:s)      & ($^{\circ}$:':'') &         &                                   &  (km s$^{-1}$)  &(Gyr)     \\
    (1)                &       (2)        &            (3)     & (4)    &          (5)                     & (6)                      &(7)        \\
\hline \hline
VCC 21            &12:10:23.14&+10:11:18.9& dE(di,bc)& 0.94 $\pm$  0.34  &   23.7 $\pm$   4.9 & 0.78  $^{+ 5.21}_{- 0.52}$\\
VCC 308          &12:18:50.77&+07:51:41.3& dE(di,bc)&  0.96 $\pm$  0.27  &  33.3 $\pm$   1.2 &2.63 $^{+ 4.72}_{- 1.71}$\\
VCC 397          &12:20:12.25&+06:37:23.6& dE(di)     &2.48 $\pm$  0.57  &   39.4 $\pm$   1.4 &1.60  $^{+ 3.21}_{- 0.85}$\\
VCC 523          &12:22:04.13&+12:47:15.1& dE(di)     &1.50 $\pm$  0.22  &   46.8 $\pm$   0.9 &3.30  $^{+ 3.84}_{- 2.00}$\\
VCC 856          &12:25:57.93&+10:03:13.8& dE(di)    & 1.04 $\pm$  0.23  &   34.0 $\pm$   3.2 &5.90   $^{+ 6.76}_{- 3.90}$\\
VCC 917          &12:26:32.40&+13:34:43.8& dE         &0.83 $\pm$  0.29  &    36.9 $\pm$   0.9 &7.42    $^{+ 9.49}_{- 5.00}$\\
VCC 990          &12:27:16.91&+16:01:28.4& dE(di)    & 0.90 $\pm$  0.06  &  44.0  $\pm$  1.4 &11.71  $^{+ 6.76}_{- 6.22}$\\
VCC 1087        &12:28:17.88&+11:47:23.7& dE         & 0.23 $\pm$  0.21  &  49.7 $\pm$   1.1 &7.32   $^{+ 9.37}_{- 5.43}$\\
VCC 1122        &12:28:41.74&+12:54:57.3& dE         & 0.47 $\pm$  0.21  &  38.6 $\pm$   0.9 &8.01   $^{+ 8.32}_{- 5.72}$\\
VCC 1183        &12:29:22.49&+11:26:01.8& dE(di)    &  0.46 $\pm$  0.13  & 42.4 $\pm$   2.2 & 3.48  $^{+ 2.54}_{- 1.06}$\\
VCC 1261        &12:30:10.35&+10:46:46.3& dE         & 0.35 $\pm$  0.13  &  50.5 $\pm$   1.3 &3.74   $^{+ 2.35}_{- 2.00}$\\
VCC 1431        &12:32:23.39&+11:15:47.4& dE         & 0.13 $\pm$  0.07  &  54.1 $\pm$  1.6 &16.14  $^{+ 9.52}_{- 8.07}$\\
VCC 1549        &12:34:14.85&+11:04:18.1& dE         & 0.31 $\pm$  0.13  &  41.1 $\pm$  2.5 &11.55  $^{+21.83}_{- 2.41}$\\
VCC 1695        &12:36:54.79&+12:31:12.3& dE(di)    &  0.94 $\pm$  0.21  & 35.1 $\pm$   1.5 &  2.75  $^{+ 4.44}_{- 1.25}$\\
VCC 1861        &12:40:58.60&+11:11:04.1& dE         & 0.15 $\pm$  0.11  &  42.2 $\pm$   1.0 & 8.47   $^{+11.52}_{- 6.31}$\\
VCC 1910        &12:42:08.68&+11:45:15.9& dE(di)    &  0.38 $\pm$  0.15  & 36.1 $\pm$   1.5 & 7.58  $^{+ 9.57}_{- 5.51}$\\
VCC 1912        &12:42:09.12&+12:35:48.8& dE(bc)   &   0.52 $\pm$  0.14  &38.6 $\pm$   1.6 &  1.35 $^{+ 2.00}_{- 0.69}$\\
VCC 1947        &12:42:56.36&+03:40:35.6& dE(di)    &  1.14 $\pm$  0.09  &49.4 $\pm$   1.3 & 2.99  $^{+ 3.68}_{- 0.89}$\\
\hline
\end{tabular}}
\end{center}
NOTES: Column 4: morphological type classification according to \citet{Lisk06b} and \citet{Lisk06a}: dE(di) indicates dwarf ellipticals with a certain, probable or possible underlying disk (i.e. showing spiral arms, edge-on disks, and$/$or a  bar) or other structures (such as irregular central features (VCC21)); dE(bc) refers to galaxies with a blue centre; dE to galaxies with no evident underlying structure. VCC1947 was not in Lisker et al. sample, therefore we classified it attending to its boxyness/diskyness, as described in Paper I. Column 5: Anisotropy parameter corrected for inclination from Paper I. Column 6: Velocity dispersion measured collapsing the spectra up to the $R_{\rm SMA}$. Column 7: Luminosity weighted ages of the stellar population from \citet{Mich08}.
\end{table*}

The sample analysed in this work (described in detail in Paper I and presented in Table \ref{sample}) consists of 18 Virgo galaxies with ${\rm M}_r> -16$ classified as dE or dS0  in the Virgo Cluster Catalog (VCC) by \citet{Bing85}. They were selected to have SDSS imaging and to be within the GALEX Medium Imaging Survey (MIS) fields \citep{Boselli05,Boselli11}, yielding a measured UV magnitude or an upper limit.

The three field dEs presented in Paper I are not included in this analysis, because the distances available show errors above 1 Mpc (using NED), reaching 14 Mpc in the case of NGC3073 \citep{SBF01}, which means errors up to $\sim 43 \%$ while for the rest of the objects the errors are below 0.5 Mpc ($0.9-3.7 \%$).

\subsection{Long-slit optical spectroscopy}

The medium resolution ($R\simeq3800$) long-slit spectroscopic observations were carried out in three different observing runs. Two out of the three runs were conducted at the 4.2m William Herschel Telescope (WHT) with the Intermediate dispersion Spectrograph and Imaging System (ISIS) double-arm spectrograph and one run at the 2.5m Isaac Newton Telescope (INT) with the Intermediate Dispersion Spectrograph (IDS). The wavelength range coverage is from 3500 \AA~ to 8950 \AA~ at the WHT, with a gap between 5000-5600  \AA~ in one of the observing campaigns, and 4600-5960 \AA~ at the INT. The resolution obtained is 1.6 \AA~ (40 km s$^{-1}$) Full Width at Half Maximum (FWHM) and 3.2 \AA~  (58 km s$^{-1}$) FWHM in, respectively, the blue and red arms of ISIS, and 1.8 \AA~ (46 km s$^{-1}$) FWHM with IDS.

The reduction has been performed through standard procedures for long-slit spectroscopy in the optical range using \reduceme~ \citep{Car99}, a package specially focused on the parallel treatment of errors. 
An accurate description of the spectroscopic data acquisition and reduction is given in the Paper I.

\subsection{Near-infrared photometry} \label{K-band}

The photometric data were obtained in the $H$ (1.65$\mu$m) or $K$ (2.2$\mu$m) bands in 6 different observing campaigns. Two out of the 6 runs were carried out at the 3.6m Telescopio Nazionale Galileo (TNG) with the Near Infrared Camera Spectrometer (NICS), three at the 4.2m WHT with Long-slit Intermediate Resolution Infrared Spectrograph (LIRIS) NIR imager and one at the 2.5m Nordic Optical Telescope (NOT) using the Nordic Optical Telescope near infrared Camera (NOTCam). All the instruments provided a resolution of 0.25$"/$pixel. The observations were carried out applying a dithering technique over 4, 8 or 9 positions on the square grid of the CCD in order to construct sky frames. Since the instruments have a 4.2$'$ $\times$ 4.2$'$ field of view and the galaxies are small ($R_{e,K} \lesssim 20 "$), this is sufficient sky coverage.

Individual exposure times varied from 10 seconds to 30 seconds in order to prevent saturation. Total integration times varied from 10 minutes up to over an hour depending on the faintness of the galaxy. All the images have been reduced using the package SNAP ({\it Speedy Near-Infrared data Automatic Pipeline, } (www.arcetri.astro.it/$\sim$filippo/snap/). This software was explicitly written for the TNG telescope's NICS instrument, but, since many important characteristics of the NICS, LIRIS and NOTCam are similar, we modified SNAP with certain scripts to successfully reduce LIRIS and NOTCam data. The photometric calibration of the data was performed using stars present in the field of view of our instruments in comparison with the same stars in the surveys of UKIDSS \citep[UKIRT Infrared Deep Sky Survey,][]{UKIDDS} or 2MASS \citep[2 Micron All Sky Survey,][]{2MASS}, when not available in UKIDSS. For those galaxies without nearby stars present in the field of view of the instruments, we have made the relative flux calibration by comparing the same galaxy in our observations with one of the surveys mentioned above.

8 out of the 18 galaxies were not observed within these campaigns, therefore we use the $H$ band images from the GOLDMine database \citep{Gavazzi01,GOLDMine}. These images come from the near-infrared imaging camera SofI (NTT telescope, ESO-Chile) and for VCC1122 from Arnica (TNG-La Palma). These images were recalibrated using UKIDSS/2MASS with the same procedures applied to our observations in order to get the most coherent sample possible.

\section{Photometric parameters}\label{FP_phot}

\begin{table*}
\begin{center}
\caption[$B$ and $K$ band photometric parameters of our sample of dEs in the Virgo cluster.]{Photometric parameters for the sample of dEs. \label{t1_FP}}
\resizebox{18cm}{!} {
\begin{tabular}{|c|c|c|c|c|c|c|c|c|c|c|}
\hline \hline
Galaxy     & ${\rm M}_V$   & $\epsilon_{V}$ & $R_{{\rm SMA},V}$ & $<\mu_{e,V}>$ & ${\rm M}_K$   & $\epsilon_{K}$ & $R_{{\rm SMA},K}$ & $<\mu_{e,K}>$ & ($V-K$)$_e$ &Ref.\\
                & (mag)    &                         &  (arcsec)  &(mag arcsec$^{-2}$)&  (mag) &                        &  (arcsec)  & (mag arcsec$^{-2}$) & mag & \\
      (1)     &    (2)      &        (3)              &   (4)        &     (5)                      &    (6)    &     (7)               &    (8)        &         (9)                    & (10) & (11)\\
\hline \hline
VCC 21            &  -16.75$\pm$0.03 &  0.36$\pm$  0.03& 13.86$\pm$0.17& 21.59$\pm$0.06 &-18.47$\pm$0.05& 0.35 $\pm$ 0.02  & 10.80$\pm$0.38  & 18.81$\pm$0.10&2.42$\pm$0.06 & 2 \\
VCC 308          &  -17.65$\pm$0.05 &  0.04$\pm$  0.03& 19.16$\pm$0.07& 21.79$\pm$0.04 &-19.79$\pm$0.07&  0.06$\pm$ 0.03 & 16.70$\pm$0.05 & 18.83$\pm$0.06&2.79$\pm$0.06 & 2\\
VCC 397          &  -16.44$\pm$0.05 &  0.33$\pm$  0.03& 13.56$\pm$0.14& 21.86$\pm$0.05 &-18.96$\pm$0.07&  0.37$\pm$  0.04 & 13.05$\pm$0.10 & 18.77$\pm$0.09&3.08$\pm$0.05 & 1\\
VCC 523          &  -18.23$\pm$0.03 &  0.25$\pm$  0.01& 26.73$\pm$0.46& 21.70$\pm$0.05 &-20.59$\pm$0.23&  0.27$\pm$  0.02 & 17.34$\pm$0.01 & 18.30$\pm$0.23&2.91$\pm$0.23 & 1\\
VCC 856          &  -17.45$\pm$0.06 &  0.08$\pm$  0.03& 16.21$\pm$0.18& 21.63$\pm$0.05 &-19.67$\pm$0.08&  0.11$\pm$  0.05 & 14.15$\pm$0.21  & 18.68$\pm$0.09&2.86$\pm$0.05 & 2\\
VCC 917          &  -16.26$\pm$0.03 &  0.41$\pm$  0.02&  9.68$\pm$0.07&  21.22$\pm$0.05 &-18.42$\pm$0.06& 0.37 $\pm$ 0.07  & 8.61$\pm$0.07  & 18.49$\pm$0.13&2.80$\pm$0.06 & 2\\
VCC 990          &  -17.13$\pm$0.03 &  0.34$\pm$  0.02&  9.88$\pm$0.06&  20.52$\pm$0.04 &-19.58$\pm$0.06& 0.36 $\pm$ 0.04  & 10.49$\pm$0.05 & 18.17$\pm$0.09&2.92$\pm$0.06 & 1\\
VCC 1087        &  -17.97$\pm$0.06 &  0.28$\pm$  0.03& 27.02$\pm$0.29& 21.94$\pm$0.06 &-20.55$\pm$0.16&   0.32$\pm$  0.04& 17.47$\pm$0.09 & 18.57$\pm$0.16&2.97$\pm$0.15 & 1\\
VCC 1122        &  -16.86$\pm$0.03 &  0.50$\pm$  0.04& 14.26$\pm$0.14& 21.27$\pm$0.10 &-18.98$\pm$0.05&   0.55$\pm$  0.08& 11.81$\pm$0.18 & 18.27$\pm$0.20&2.82$\pm$0.05 & 2\\
VCC 1183        &  -17.55$\pm$0.03 &  0.22$\pm$  0.12& 21.85$\pm$0.28& 21.99$\pm$0.17 &-19.98$\pm$0.05&   0.31$\pm$  0.09& 19.37$\pm$0.42 & 18.80$\pm$0.16&3.09$\pm$0.05 & 2\\
VCC 1261        &  -18.38$\pm$0.06 &  0.37$\pm$  0.05& 23.76$\pm$0.21& 21.28$\pm$0.08 &-20.94$\pm$0.15&   0.41$\pm$  0.06& 20.37$\pm$0.10 & 18.32$\pm$0.18&2.80$\pm$0.14 & 1\\
VCC 1431        &  -17.28$\pm$0.06 &  0.03$\pm$  0.01& 10.31$\pm$0.05& 20.78$\pm$0.03 &-20.26$\pm$0.16&   0.02$\pm$  0.02& 9.91$\pm$0.08 & 17.73$\pm$0.15&3.11$\pm$0.15 & 1\\
VCC 1549        &  -16.96$\pm$0.03 &  0.16$\pm$  0.01& 13.09$\pm$0.08& 21.55$\pm$0.03 &-19.43$\pm$0.06&   0.19$\pm$  0.01& 11.83$\pm$0.08 & 18.36$\pm$0.06&3.10$\pm$0.06 & 2\\
VCC 1695        &  -17.33$\pm$0.08 &  0.22$\pm$  0.05& 27.61$\pm$0.73& 22.69$\pm$0.09 &-19.11$\pm$0.13&   0.16$\pm$  0.07& 18.43$\pm$0.06 & 19.24$\pm$0.14&2.86$\pm$0.10 & 2\\
VCC 1861        &  -17.55$\pm$0.06 &  0.04$\pm$  0.02& 21.57$\pm$0.28& 22.12$\pm$0.04 &-20.60$\pm$0.09&   0.01$\pm$  0.01& 20.11$\pm$0.08 & 18.94$\pm$0.07&3.21$\pm$0.08 & 1\\
VCC 1910        &  -17.43$\pm$0.06 &  0.14$\pm$  0.04& 14.24$\pm$0.11& 21.20$\pm$0.06 &-19.55$\pm$0.08&   0.17$\pm$  0.01& 11.92$\pm$0.04 & 18.65$\pm$0.06&2.88$\pm$0.06 & 1\\
VCC 1912        &  -17.58$\pm$0.03 &  0.54$\pm$  0.06& 23.53$\pm$0.17& 21.55$\pm$0.15 &-20.18$\pm$0.05&   0.59$\pm$  0.08& 21.14$\pm$0.58 & 18.59$\pm$0.23&2.77$\pm$0.05 & 1\\
VCC 1947        &  -17.33$\pm$0.03 &  0.23$\pm$  0.01& 11.26$\pm$0.08& 20.76$\pm$0.03 &-20.31$\pm$0.12&   0.23$\pm$  0.01& 9.56$\pm$0.05 & 17.42$\pm$0.12&3.25$\pm$0.12 & 1\\
\hline
\end{tabular}}
\end{center}
NOTES: Columns 2, 3, 4 and 5 present the absolute magnitude, the ellipticity, the half light radius along the semimajor axis, and the effective surface brightness within this effective radius in $V$ band (Johnson-Cousins), respectively. Columns 6, 7, 8 and 9 show the same parameters in $K$ band. All the magnitudes used throughout this work are referred to Vega. The transformation from this half light radius to the one used in the scaling relations, the geometric half light radius, is $R_e=R_{\rm SMA}\sqrt{1-\epsilon}$. Column 10 is the $(V-K)_e$ colour measured within an aperture with the size of the half light radius of each galaxy measured in the SDSS $g$ band. Column 11 specifies the reference of the infrared image: 1 means this work (from $H$ or $K$ band photometry), 2 means that the image was gathered from GOLDMine database \citep[$H$ band,][]{GOLDMine}. The distances assumed for each galaxy are calculated from their surface brightness fluctuations by \citet{Mei07}, or similarly using $H_0$ $=$ 73 km s$^{-1}$ Mpc$^{-1}$, see Paper I for the list of individual distances.
\end{table*}

Photometric parameters have been determined in the $K$ and $V$ bands using Vega magnitudes.
As specified in Table \ref{t1_FP}, not all the galaxies were observed in the $K$ band, for those only available in the $H$ band we have assumed a colour of $H-K=$ 0.21 \citep{Pel99}.
The $V$ band (Johnson-Cousins) is used to make comparisons with previous studies that are mostly done in the optical. The photometric parameters for our sample of dEs in this optical band have been obtained from the analysis of $g$ and $r$ bands from the Sloan Digital Sky Survey \citep[SDSS,][]{SDSS} Data Release 7 \citep[DR7,][]{SDSS_DR7}. The transformation into the $V$ band has been done following \citet{Blanton07}:
$ V=g-0.3516-0.7585[(g-r)-0.6102]$, 
where $g$ and $r$ asymptotic magnitudes have been measured as described in Section \ref{method}. To transform the AB magnitudes of SDSS photometry into Vega magnitudes we used $m_{V,Vega}=m_{V,AB}-0.02$ mag \citep{Blanton07}.

\subsection{Methodology}\label{method}

The parameters used in the scaling relations are asymptotic magnitudes, effective radii $R_e$ (geometric radius containing 50$\%$ of the total light), effective surface brightness $<\mu_e>$ (the mean surface brightness within $R_e$ ) and the ellipticity $\epsilon$. 
All these parameters, presented in Table \ref{t1_FP}, were calculated using the IRAF\footnote{IRAF is distributed by the National Optical Astronomy Observatory, which is operated by the Association of Universities for Research in Astronomy, Inc., under the cooperative agreement with the National Science Foundation.} task {\sc ellipse} and following the same procedure as described in Paper I.

The first step before obtaining any photometric parameters was to remove the stars surrounding our target galaxies to avoid their contribution to the flux in the galaxy isophotes.
Stars present in the $g$ band images were removed using the masks we obtained for the $i$ band SDSS images in Paper I. We matched the coordinates of both $g$ and $i$ bands using the IRAF task {\sc wregister}. The mask of stars for the $H/K$ band was generated  with {\sc ellipse}.
To remove the stars from the images we used the IRAF task {\sc fixpix} that interpolates the masked regions by the surrounding area. To improve the final outcome, we averaged the results of interpolating along the horizontal and vertical directions for each star.
For those galaxies that were on the edge of the FITS images or had a very bright nearby star, a more careful procedure was followed. Taking advantage of the fact that the galaxies have smooth surface brightness profiles, the {\sc bmodel} task of IRAF can be used to replace the affected areas of the galaxy by the azimuthal average of the unaffected ones.  The output from {\sc bmodel} was used only to replace a small fraction of pixels, so that the possible presence of more subtle features, such as bars or spiral arms, remain. Once very bright nearby stars were removed, the remaining stars were cleaned following the same procedure with {\sc fixpix} as described above.

The second step was to run {\sc ellipse} fixing only the center of the galaxy and leaving the rest of parameters free to measure the ellipticity of the isophotes and the position angle (PA) of their major axis in each band. The $\epsilon$ value that we assigned for each galaxy was the mean value between 3$"$ and the half light radius along the semimajor axis ($R_{\rm SMA}$), the galaxy region covered by our spectroscopic observations. The error of the ellipticity has been considered as the scatter of $\epsilon$ in this region.  For the position angle we just used the typical values in the outer parts of the galaxy (beyond 1.5$-$2 times the half light radius, region where this parameter stabilise). This step was skipped for the optical images because we used the $\epsilon$ and PA obtained in $i$ band following the same method (see Paper I).

To obtain the asymptotic magnitudes and the half light radius we run {\sc ellipse} again, with a step between consecutive isophotes of 1 pixel, and with the center of the galaxy, the ellipticity and the PA fixed to avoid overlap between consecutive isophotes. The procedure followed to obtain these parameters is described in \citet{GdP07}. We first computed the accumulated flux and the gradient in the accumulated flux (i.e., the slope of the growth curve) at each radius, considering as radius the major-axis value provided by {\sc ellipse}. Secondly, we chose a radial range where this gradient had a linear regime, which is in the outer parts of the galaxy where the growth curve becomes flat or nearly flat. Finally, we performed a linear fit to the accumulated flux as a function of the slope of the growth curve. The asymptotic magnitude of the galaxy is the y-intercept, or, equivalently, the extrapolation of the growth curve to infinity. Once the asymptotic magnitude is known, the half light radius is determined from the growth curve corresponding to the major-axis of the elliptical isophote containing 50$\%$ of the total flux, $R_{\rm SMA}$. (See Appendix \ref{app_errors} for a description of the errors in the asymptotic magnitudes).

To obtain the mean effective surface brightness within the effective radius we measured half the total flux inside the area of an ellipse with semi-major axis $R_{\rm SMA}$ 

\begin{equation}\label{mueff}
<\mu_e>=m+2.5log(2)+2.5log(\pi R_{\rm SMA}^2(1-\epsilon))
\end{equation}

\noindent with $m$ the asymptotic magnitude. The errors of the surface brightness were calculated propagating the errors of the magnitudes, $R_{\rm SMA}$ and $\epsilon$.
The half light radius we use in the scaling relations are geometric radius $R_e=R_{\rm SMA}\sqrt{1-\epsilon}$.

The colour $(V-K)_e$ is measured in a fixed aperture with radius equal to the half light radius in the SDSS $g$ band. The very different depth of the optical and near-infrared images prevents us from using infinite apertures.

\subsection{Comparison with the literature}

\begin{figure*}
\centering
\resizebox{0.25\textwidth}{!}{\includegraphics[angle=-90]{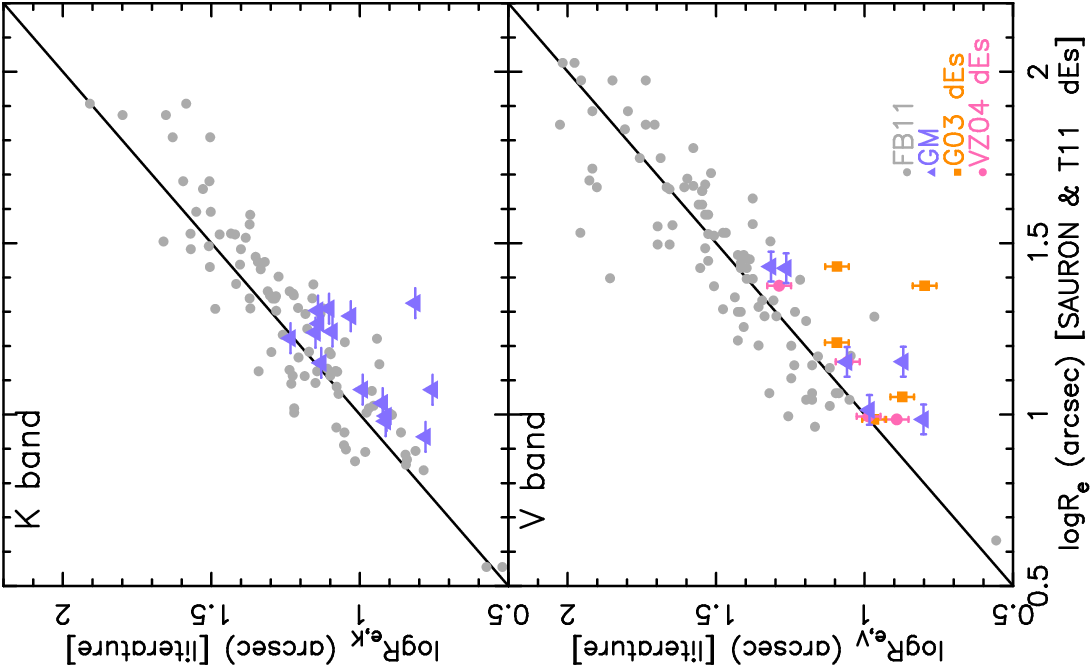}}
\resizebox{0.25\textwidth}{!}{\includegraphics[angle=-90]{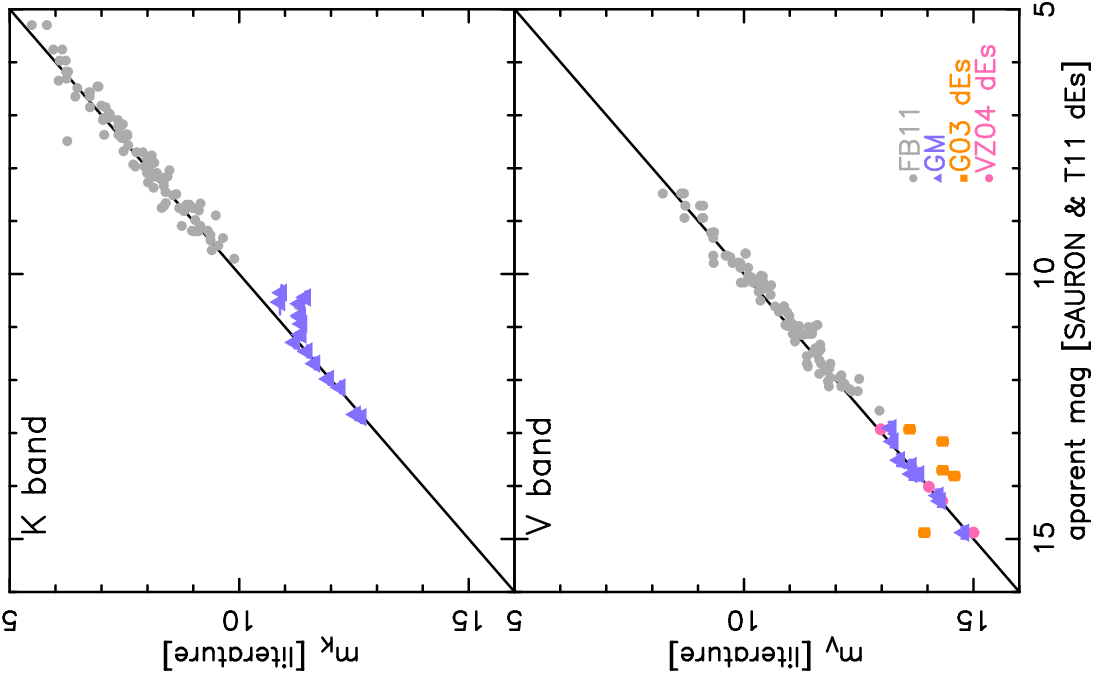}}
\resizebox{0.48\textwidth}{!}{\includegraphics[angle=-90]{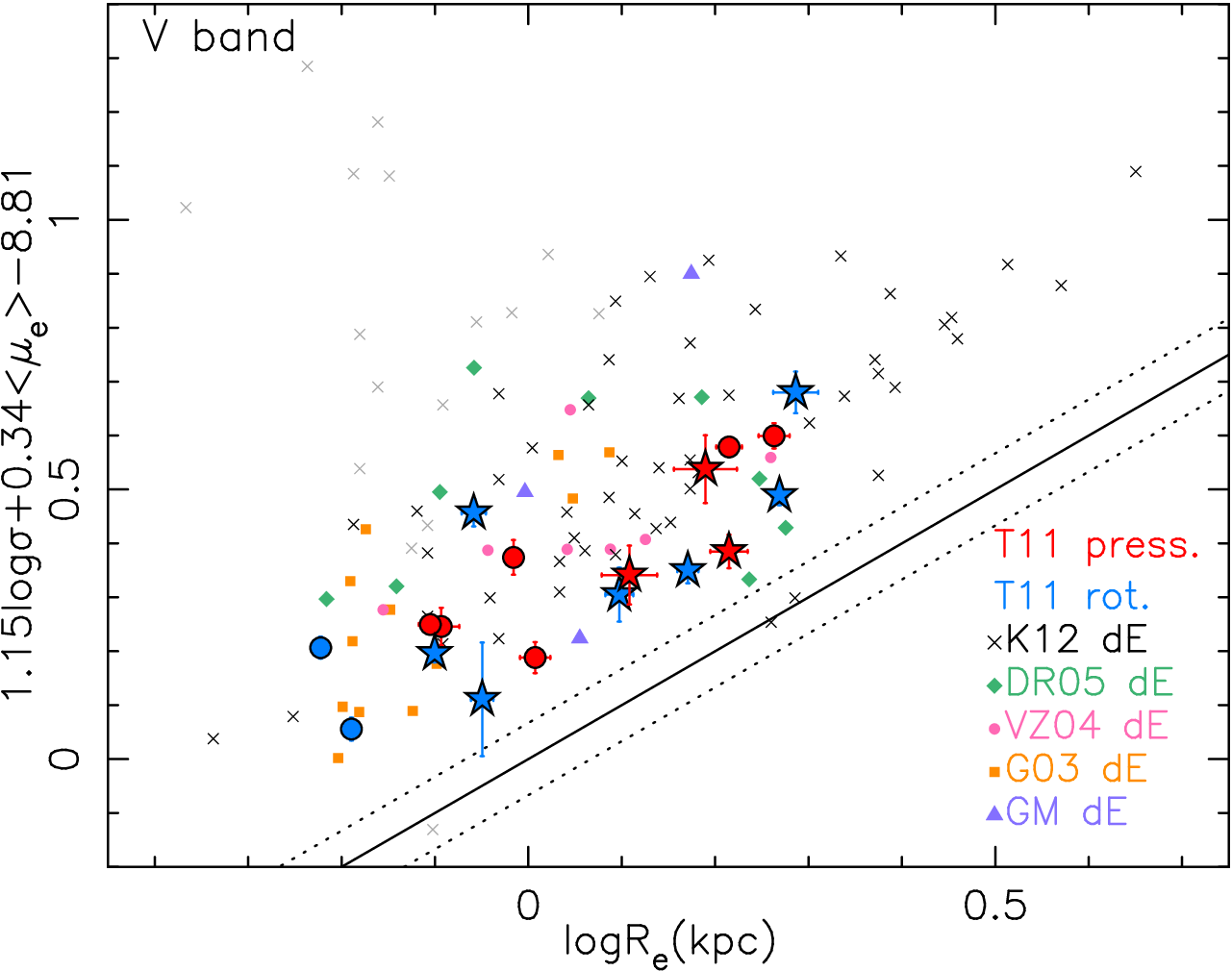}}
\caption{Comparison of the geometric effective radius (left panels), and apparent magnitudes (central panels) between different sources from the literature. In grey are shown those samples in common with the SAURON galaxies and in colours those in common with our sample of dEs. The samples in common with SAURON \citep[][]{deVaucoul91,P99,Jarrett00,Kormendy09} were analysed in \citet{FB11}. In common with our dEs (T11 dEs) are \citet[][GOLDMine database or GM hereafter, G03 and VZ04, respectively]{GOLDMine,Geha03,VZ04}. The right panel shows the location of different samples of dEs on the Fundamental Plane fitted for the SAURON galaxies by \citet{FB11} and their $1\sigma$ scatter (solid and dotted lines respectively). With respect to the left and central panels we add the Fornax sample of dEs by \citet[][DR05]{DR05} and the dEs from Coma by \citet[][K12]{Kourkchi12b}. Those Coma dEs fainter than $M_V=-17$ mag are plotted in grey.
}
              \label{reff_mag_comps}
\end{figure*}

Only few spectrophotometric studies of Virgo dEs are available in the literature. In the left and central panels of Figure \ref{reff_mag_comps} we show a comparison of the geometric half light radii and apparent magnitudes, in the $V$ and $K$ bands, for the galaxies in common with our sample of dEs. We added here the literature comparison for a sample of large galaxies (the SAURON study by \citet[][hereafter FB11]{FB11}).  Only the sample by \citet[][VZ04]{VZ04} provides geometric radii, thus we homogenized the $R_{\rm SMA}$ using the ellipticity for each system, which in both cases are provided, \citet[][the GOLDMine database, hereafter GM]{GOLDMine} and \citet[][hereafter G03]{Geha03}.

As expected we find a better agreement for the magnitudes than for the radii. When measuring magnitudes most papers extrapolate the light outwards. This method generally does not create too large uncertainties. However, when determining radii fitting e.g. S\'ersic profiles to the surface brigtness distribution the results are strongly degenerate with surface brightness and also depend on the depth of the photometric images (e.g. {\it HST} in G03 versus SDSS in T11). In any case, the scatter found for the galaxies in common with other works, except for one outlier, is similar to the one found for massive galaxies. 

The right panel of Figure \ref{reff_mag_comps} presents the Fundamental Plane for dEs in comparison to the fit for the SAURON sample of galaxies. We have added here the samples of dEs by \citet[][hereafter DR05]{DR05} and \citet[][hereafter K12]{Kourkchi12b}, Fornax and Coma dEs respectively. For the Virgo cluster galaxies we have considered the distances calculated in \citet[][HST$/$ACS survey]{Mei07} for individual galaxies when available, or the mean distance to the A or B cluster substructure also from \citet[][$16.7\pm 0.2$ and $16.4\pm 0.2$ Mpc, respectively]{Mei07}. To know in which Virgo cluster substructure the galaxies reside we have used the Virgo Cluster Catalog \citep[][VCC]{Bing85}. For Coma and Fornax cluster galaxies we have used the mean distance to the cluster by \citet[][HST$/$ACS survey, $102\pm 14$ Mpc]{Thomsen97} and  \citet[][HST$/$ACS survey, $20.0\pm 0.3$ Mpc]{Blakeslee09} respectively.

The large spread of dEs compared to our much tighter relation is likely for a large part caused by the different methods followed to measure the velocity dispersion of the galaxies. The GM sample in itself mixes velocity dispersions from a large number of different datasets that have different resolutions and are based on different procedures to extract the velocity dispersions (see GM for details). Moreover, while K12 measures central velocity dispersions, G03 and VZ04 average the $\sigma$ profile of the galaxies, and DR05 measures a luminosity weighted $\sigma$ within a circular aperture after correcting for $h_4$, a measure of the kurtosis.

Following the works where both late-type and early-type galaxies are analysed together in the kinematic scaling relations \citep[e.g.][FB11]{Zaritsky08,Zaritsky11} we use $\sigma_e$, measured after first collapsing the spectra within $R_{\rm SMA}$. As a consequence, the lines in the individual spectra are broadened not only by random motions but also by rotation, and hence it is a proxy for kinetic energy. FB11 also measures the velocity dispersion within the half light radius so that we can directly compare our results for dEs with the SAURON sample of galaxies.


\section{Scaling Relations} \label{scaling_rels}

The aim of the present paper is to make the first study of the Faber-Jackson relation and the Fundamental Plane of a sample of dEs in the near-infrared, a wavelength range insensitive to the dust extinction that measures the bulk of the stellar populations. In order to make a complete and coherent study, we also discuss the colour - $\sigma_e$ relation and include an appendix with purely photometric scaling relations that involve size, surface brightness, magnitudes and colours.

Due to the novelty of this near-infrared wavelength band for dwarf galaxies, we also study these scaling relations in the $V$ band to be able to compare with the data available in the literature (Figure \ref{reff_mag_comps}), and also to investigate the effects of the stellar populations.

To study whether dEs follow the scaling relations of Es we need a coherent and homogeneous sample of galaxies to compare with. We choose the SAURON galaxies as control sample because in FB11 they analyse all these scaling relations in ${\it Spitzer}/Irac$  $3.6 \mu{\rm m}$ band, very similar to the $K$ band, and in the optical $V$ band. In that paper they fit these relations, compare with the rest of the literature and discuss the different systematic effects and uncertainties. Thus, we will use here the same fits to analyse the position of dEs with respect to them.
The SAURON data and fits at $3.6\mu{\rm m}$ have been transformed to the $K$ band assuming that the geometric half light radius in both bands is the same, as shown in FB11, and a colour of $K-[3.6]=0.1$, as measured also in that paper.

The SAURON distinction between slow and fast rotators \citep{Em07} is different from our definition of pressure and rotationally supported dEs ($(v_{max}/\sigma)^* \gtrless 0.8$), therefore we will not use here any division for the SAURON galaxies and we will treat all of them as early-types.

The symbols that are used throughout this work are the following: dark grey dots and light grey diamonds are the early-type and Sa SAURON galaxies respectively (FB11). The red and blue symbols are the dEs presented in Paper I (T11 in the diagrams), with pressure supported dEs ($(v_{max}/\sigma)^* < 0.8$) in red and rotationally supported dEs ($(v_{max}/\sigma)^* > 0.8$) in blue (see Table \ref{sample} and Paper I). The shape of the symbol for our sample of dEs indicates the age of the stellar populations as determined by  \citet{Mich08}. Galaxies older than 7 Gyr are indicated with circles, and galaxies younger than 7 Gyr with asterisks. In general, young dEs (asterisks) are rotationally supported (blue) while the old ones (circles) are pressure supported (red). Grey open triangles and squares are Milky Way and M31 dwarf spheroidal galaxies \citep{Wolf10,Brasseur11,Tollerud12}.

\subsection{Colour$- \sigma_{\rm e}$ relation}\label{col_rels}

\begin{figure}
\centering
\resizebox{0.45\textwidth}{!}{\includegraphics[angle=-90]{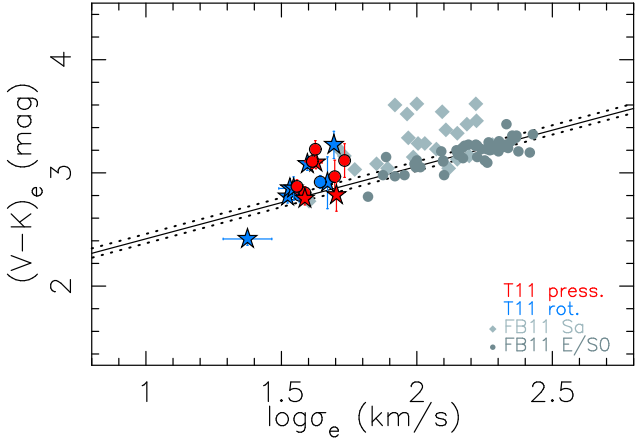}}
\caption{Colour-$\sigma_e$ diagram. The $(V-K)_e$ colour has been measured in a fixed aperture equal to the half light radius of each galaxy. The new dE sample presented in this work (T11) is shown with the red (pressure supported) and blue (rotationally supported) symbols. Stars indicate dEs younger than 7 Gyr while circles older than 7 Gyr. The more massive hot stellar systems are the SAURON sample \citep[][FB11]{FB11}. The dark grey dots and light grey diamonds are the early-types and Sa galaxies respectively. 
The solid black line is the fit for E/S0 done by FB11 (Sa galaxies not included in the fit). The dotted lines denote the $1\sigma$ scatter in the relation for the fitted galaxies.}
              \label{Col-sig}
\end{figure}

Figure \ref{Col-sig} shows the $(V-K)_e$ colour versus $\sigma_e$ - both measured within the $R_e$ - of the SAURON galaxies and T11 Virgo dEs.
The $(V-K)_e$ colour, due to its large baseline, helps us to distinguish between different stellar populations. The dEs spread $\sim1$ mag in this colour, which is the direct consequence of their range in age and metallicity \citep{Mich08}. This correlation between the mean velocity dispersion of the stellar population of the galaxy and its colour was previously found for massive galaxies \citep[e.g.][FB11]{Bernardi03,Grav09}, and it is in agreement with a {\it downsizing} star formation history. In this scenario the mass of the galaxies is the parameter that drives galaxy evolution, being low-mass galaxies those with star formation histories more extended in time than massive galaxies \citep{Cowie96,Gavazzi96,Boselli01,Caldwell,Nel05,Bundy06}. 

Some of the dEs seem to be above the relation for Es. In FB11 it is argued that the Sa galaxies that are above this relation are in fact deviating from it because of their dust content. This argument is not valid for the dEs, because of their lack of dust (usually associated to cold or warm gas, but no emission lines are seen in this sample of dEs). It is not a metallicity effect, because looking at how the metallicity affects this colour (FB11), it is seen that it goes in the opposite direction (see also Figure \ref{CMD}). 
A similar effect has been seen by \citet[Fig. 13]{Hammer10} for compact ellipticals in the Coma cluster in an optical colour. To understand this effect more and better quality data is needed.

\begin{table*}
\begin{center}
\caption{Averaged perpendicular distance and $rms$ $(d_{\bot} \pm $rms$)$ of our sample of dEs to the Faber-Jackson and Fundamental Plane relations of massive hot stellar systems (with and without the stellar mass-to-light ratio $\Upsilon_*$). $\Delta_{1\sigma}=d_{\bot}/\Delta_{\rm fit}$ quantifies these distances as a function of the $1\sigma$ scatter of Es in these relations. If $\Delta_{1\sigma}>1.0$ then the dEs are outside the fit of Es in more than $1\sigma$ scatter. \label{dists_rels}}
\begin{tabular}{|lccccccc|}
\hline \hline
                    &  Band & $d_{\bot}$ (FJ) &  $\Delta_{1\sigma}$ (FJ) & $d_{\bot}$ (FP) & $\Delta_{1\sigma}$ (FP) & $d_{\bot}$ (FP,$\Upsilon_*$)  & $\Delta_{1\sigma}$ (FP,$\Upsilon_*$) \\
\hline \hline
All dEs    &  $ K$   &   0.11 $\pm$ 0.09   &   0.99 $\pm$ 0.81 & 0.16 $\pm$ 0.06   & 3.30 $\pm$ 1.23 &   0.10 $\pm$ 0.07  &   2.45 $\pm$ 1.71 \\
Rot. supported dEs    &  $ K$   &   0.13 $\pm$ 0.10   &   1.19 $\pm$ 0.87 &  0.16 $\pm$ 0.08   & 3.25 $\pm$ 1.64 &   0.11 $\pm$ 0.09  &  2.69 $\pm$ 2.13 \\
Press. supported dEs    &  $ K$   &   0.09 $\pm$ 0.08   &  0.79 $\pm$ 0.75 & 0.17 $\pm$ 0.04   &  3.34 $\pm$ 0.75 & 0.09 $\pm$ 0.05  &  2.20 $\pm$ 1.26 \\
\hline
All dEs    &  $ V$   &  0.08 $\pm$ 0.10   & 0.54 $\pm$ 0.69 & 0.21 $\pm$ 0.07  & 4.45 $\pm$ 1.53 & 0.09 $\pm$ 0.08 &  2.01 $\pm$ 1.68 \\
Rot. supported dEs    &  $ V$   & 0.08 $\pm$ 0.13    &   0.55 $\pm$ 0.85 & 0.21 $\pm$ 0.09   & 4.38 $\pm$ 1.86 &  0.10 $\pm$ 0.09  & 2.10 $\pm$ 2.05\\
Press. supported dEs    &  $V$   &   0.08 $\pm$ 0.08   & 0.54 $\pm$ 0.55 &  0.21 $\pm$ 0.06   & 4.51 $\pm$ 1.25 &  0.09 $\pm$ 0.06  &  1.91 $\pm$ 1.32\\
\hline
\end{tabular}
\end{center}
NOTES: The $1\sigma$ scatter of the Faber-Jackson and Fundamental Plane fits reported by FB11 have been transformed to be measured perpendicularly to the fits.
\end{table*}

\subsection{Faber-Jackson relation}\label{sec_FJ}

\begin{figure}
\centering
\resizebox{0.48\textwidth}{!}{\includegraphics[angle=-90]{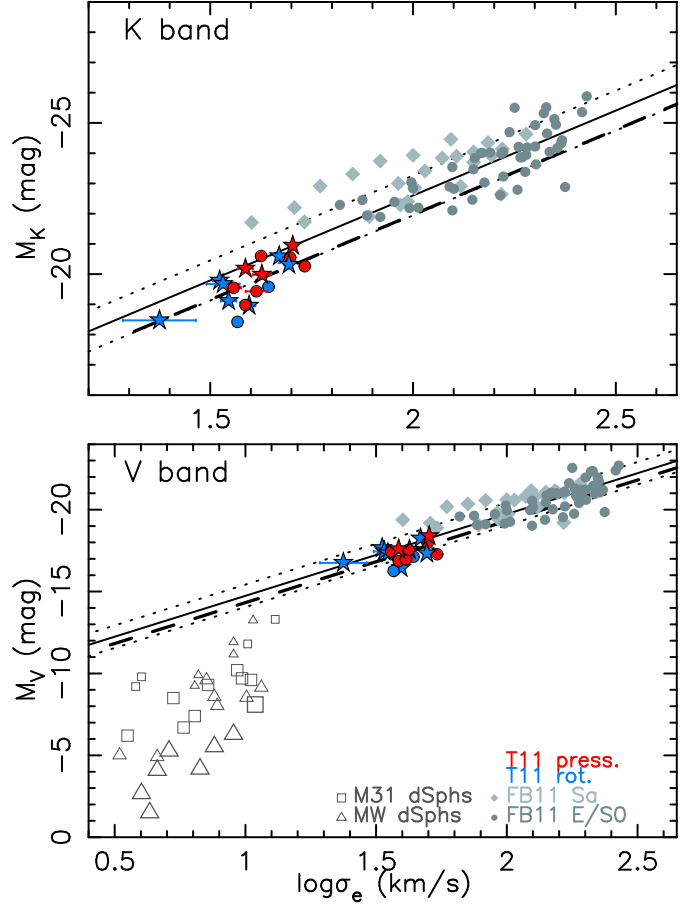}}
\caption{Faber-Jackson relation in the $K$ (upper panel) and $V$ (lower panel) bands. Solid and dotted lines are the fit and $1\sigma$ scatter for E/S0 by FB11. Symbols as in Figure \ref{Col-sig}. The black dashed line is the average perpendicular distance of T11 dEs with respect to this fit. The grey open symbols in the $V$ band are the dwarf spheroidal galaxies (dSphs) from the Milky Way (MW) and Andromeda (M31). The size of the symbol is related to the dynamical mass-to-light ratio at the half light radius of the galaxy following the values by \citet{Wolf10,Tollerud12}. The smallest symbols are for dSphs with ${\rm M_{dyn,1/2}/L_{1/2}}<50 {\rm M_{\odot}/L_{\odot,V}}$, and the biggest for ${\rm M_{dyn,1/2}/L_{1/2}}>500 {\rm M_{\odot}/L_{\odot,V}}$.}
              \label{FJ}
\end{figure}

Figure \ref{FJ} shows the Faber-Jackson relation in the $K$ and $V$ bands. We can compare the position of the dEs with respect to massive early-type systems for which the fits are shown as black solid lines (FB11). The dashed grey line is the average distance of our sample of dEs with respect to these fits. In the $K$ band we find that this distance is consistent with the $1\sigma$ scatter of the fitted galaxies (dotted lines). However, in the $V$ band this distance is even smaller than the scatter of the SAURON elliptical galaxies. Thus, although in both cases their distance is consistent with the scatter of the early-type galaxies, the dEs are grouped below the fit. When compared with the position of dSphs, the dEs seem to be the population where the Faber-Jackson relation begins to bend downwards. The perpendicular distance of the dSphs with respect to the fit of E/S0 galaxies is related to their dynamical mass-to-light ratio, thus to their dark matter fraction, which then suggests that dEs have a larger amount of dark matter than E/S0 galaxies.

Table \ref{dists_rels} shows the average perpendicular distance, and its $rms$, that the dEs have with respect to the fits of massive Es ($d_{\bot}$). We have also divided these distances by the $1\sigma$ scatter ($\Delta_{\rm fit}$) of the SAURON early-type galaxies ($\Delta_{1\sigma}=d_{\bot}/\Delta_{\rm fit}$). This quantity indicates whether the location of the dEs is statistically different from the distribution of more massive systems ($\Delta_{1\sigma}>1.0$). In the case of the Faber-Jackson relation, rotationally and pressure supported dEs are indistinguishable, both show the same  average distance with respect to the massive ellipticals within the errors. However, the intrinsic scatter ($rms$ of $d_{\bot}$) of pressure supported dEs is systematically smaller than the values found for the rotationally supported ones.

\subsection{Fundamental Plane}\label{sec_FP}

The Fundamental Plane (FP), originally defined by \citet{Djor87} and \citet{Dress87}, is the flat surface that best fits massive early-type galaxies in the three-dimension space defined by the effective surface brightness, the velocity dispersion and the effective radius. The edge-on view of the FP is widely expressed as:

\begin{equation}
\log R_e= \alpha ~  \log \sigma_e + \beta ~ <\mu_e> + \gamma
\end{equation}

\noindent 
where $\alpha$, $\beta$ and $\gamma$ are the coefficients that minimise the scatter of the FP. 

\begin{figure*}
\centering
\resizebox{1.\textwidth}{!}{\includegraphics[angle=-90]{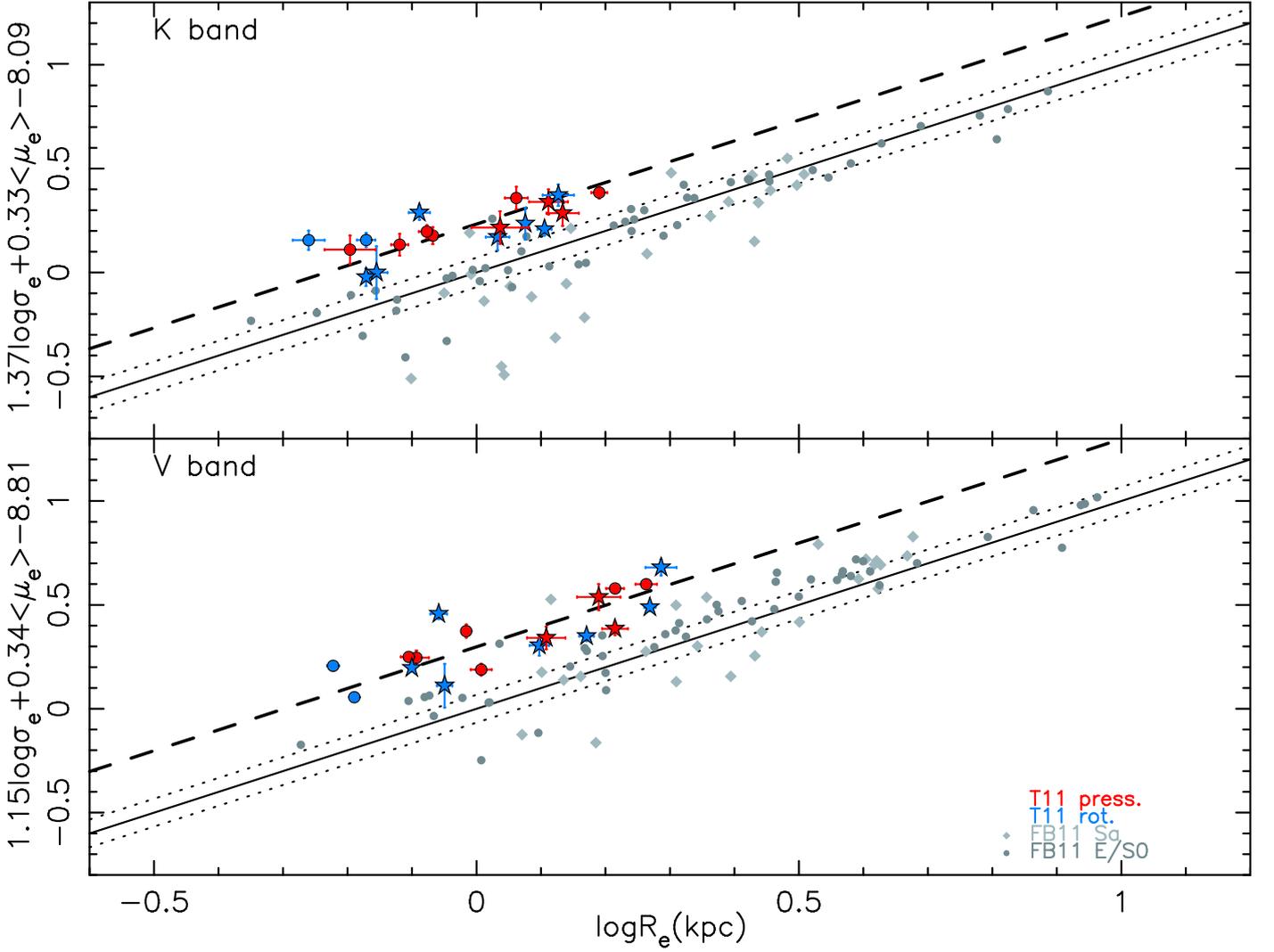}}
\caption{Fundamental plane. Symbols as in Figure \ref{Col-sig}. The continuum line show the fit by \citet{FB11} and the dotted lines present the $1\sigma$ scatter for all the massive early-type galaxies plotted. The black dashed line is the average perpendicular distance of our dEs to the Fundamental Plane.}
              \label{FP}
\end{figure*}

Figure \ref{FP} presents this edge-on view of the FP in the $K$ and $V$ bands. The fits (solid lines) and $1\sigma$ scatter (dotted lines) are those performed by FB11 for massive early-type galaxies. In the radius range in which Es and dEs overlap, dEs, instead of following the same plane as Es, lie above it. The grey dashed lines show the average perpendicular distance of the dEs with respect to the FP of E/S0 galaxies, and Table \ref{dists_rels} quantifies it as a function of the $1\sigma$ scatter of the FP for massive early-type systems. This distance is a bit more than $3\sigma$ in the $K$ band, and about $4\sigma$ in the $V$ band. In addition, if we look at the distance of pressure and rotationally supported dEs individually (Table \ref{dists_rels}), we see that, although the intrinsic scatter of pressure supported dEs ($rms$ of $d_{\bot}$) is significantly smaller than that of rotationally supported dwarfs, both populations do not show, within the errors, any difference in their offset with respect to the plane.

\subsection{Fundamental Plane in the $\kappa$-space}

\begin{figure*}
\centering
\resizebox{0.495\textwidth}{!}{\includegraphics[angle=-90]{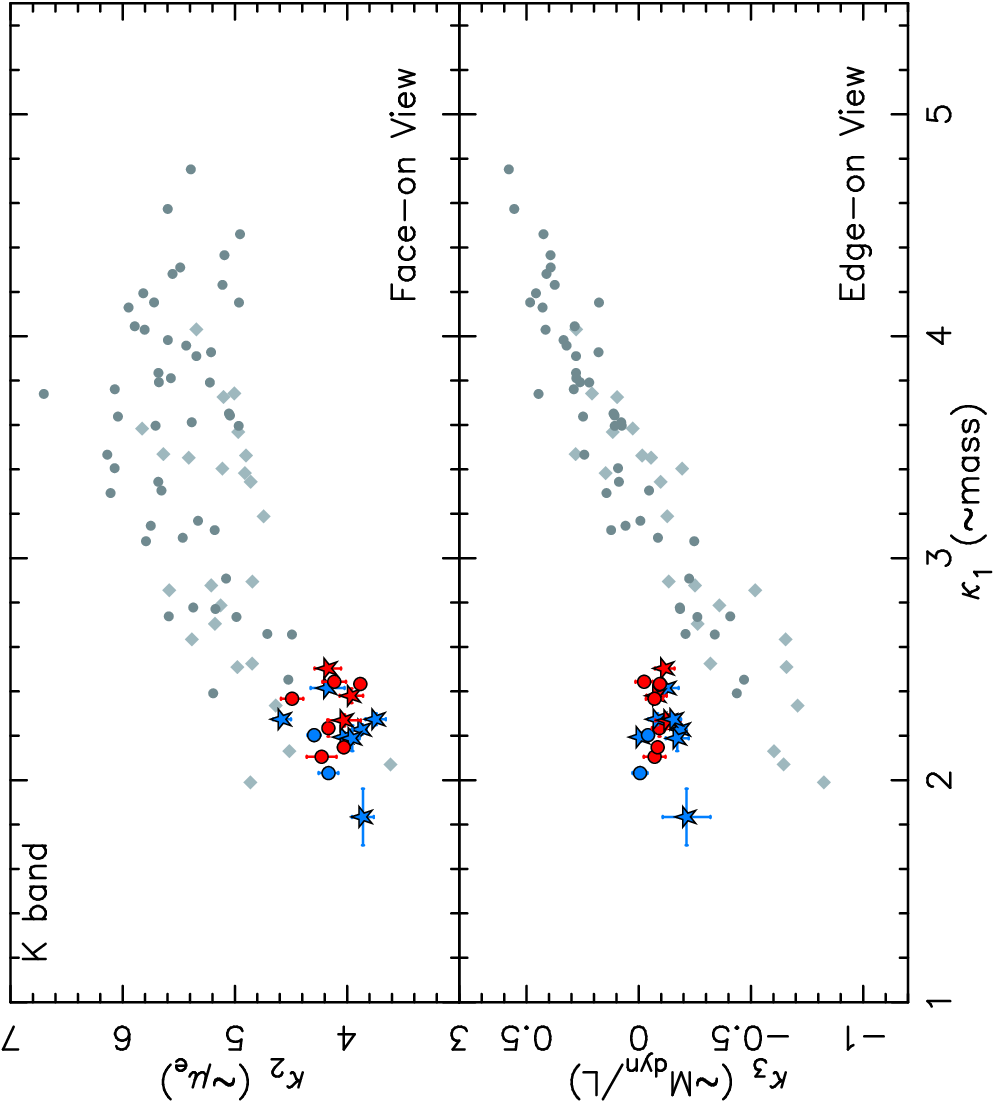}}
\resizebox{0.495\textwidth}{!}{\includegraphics[angle=-90]{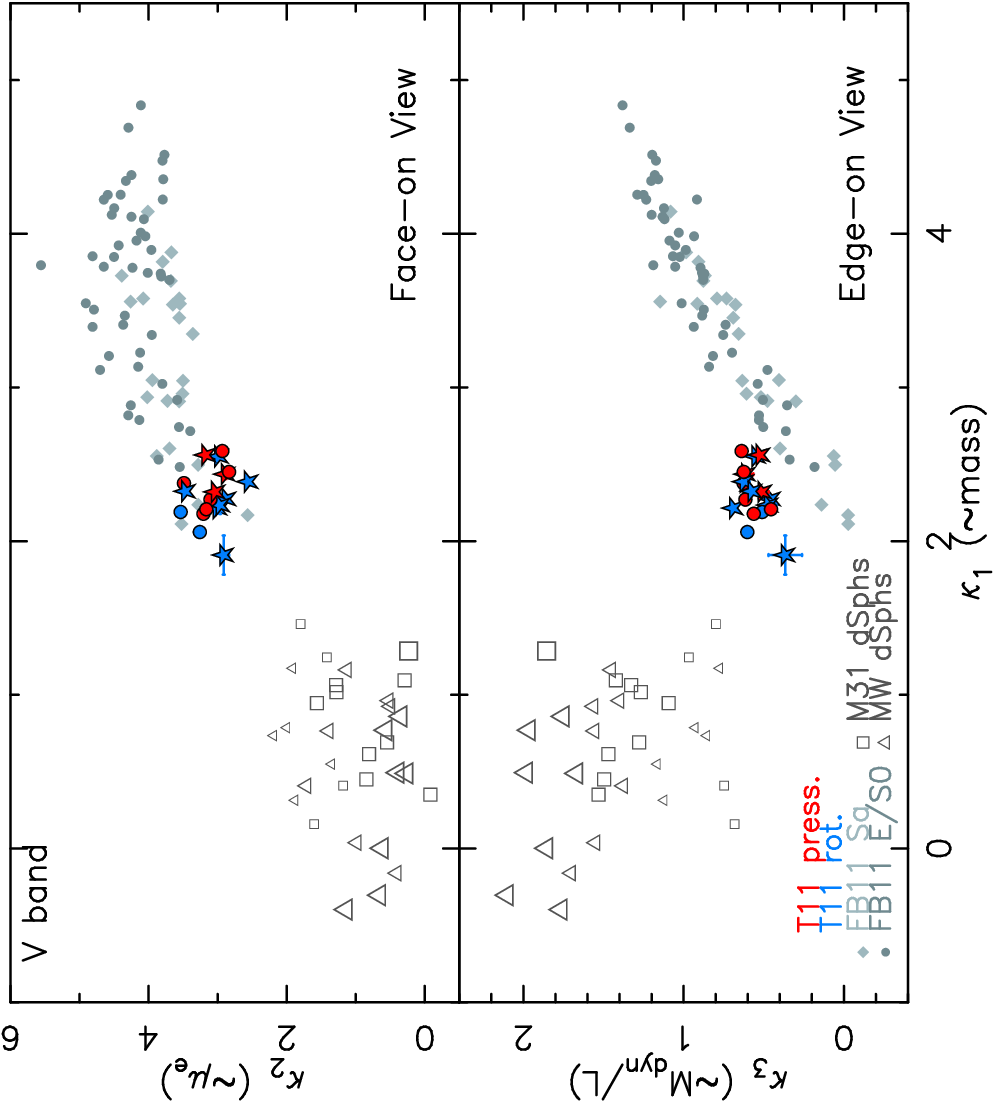}}
\caption{Face-on (upper panels) and edge-on (lower panels) views of the Fundamental Plane in the $\kappa$-space. Symbols as in Figure \ref{FJ}.}
              \label{FP_kappa}
\end{figure*}

\citet{Bender92} re-cast the FP into a $\kappa$-space in which the axes are proxies for mass ($\kappa_1$), effective surface brightness ($\kappa_2$), and dynamical mass-to-light ratio ($\kappa_3$).
Figure \ref{FP_kappa} shows the edge-on and face-on projections of this $\kappa$-space. In the $V$ band we have added dSph galaxies from the Milky Way and Andromeda \citep{Wolf10,Brasseur11,Tollerud12}.

In the face-on view, dEs continue the sequence of Es and Sa galaxies to lower masses, and this sequence can be extended down to dSphs. However, in the edge-on view dEs do not follow the trend of more massive systems, but bend towards the dSph galaxies. 
We see that for the same mass, dEs have a larger $M_{\rm dyn}/L$ than massive early-type galaxies. Moreover, the overall shape of this edge-on view is the U-shape identified in Paper I when the dynamical mass-to-light ratio is plotted versus the absolute magnitude of these systems \citep[see also][]{Zaritsky06,Wolf10}. This U-shape plot has been used as a clear proof of the large dark matter content of dSphs. Those with low amounts of dark matter appear near the dEs (the smallest symbols in Figure \ref{FP_kappa}), while those highly dark matter dominated are further away (large symbols). This suggests that the $M_{\rm dyn}/L$ is the reason why dEs are shifted above the FP of massive objects.

G03 also analysed this $\kappa$-space in the $V$ band, but, due to the lack of a sample of early-types with a large range in mass, they found that dEs were isolated in an area distinctly different from other stellar systems. Now, we do not see any empty region between dEs and Es.

\subsection{Corrections to the FP: dark matter content of dEs}

\begin{figure*}
\centering
\resizebox{1.\textwidth}{!}{\includegraphics[angle=-90]{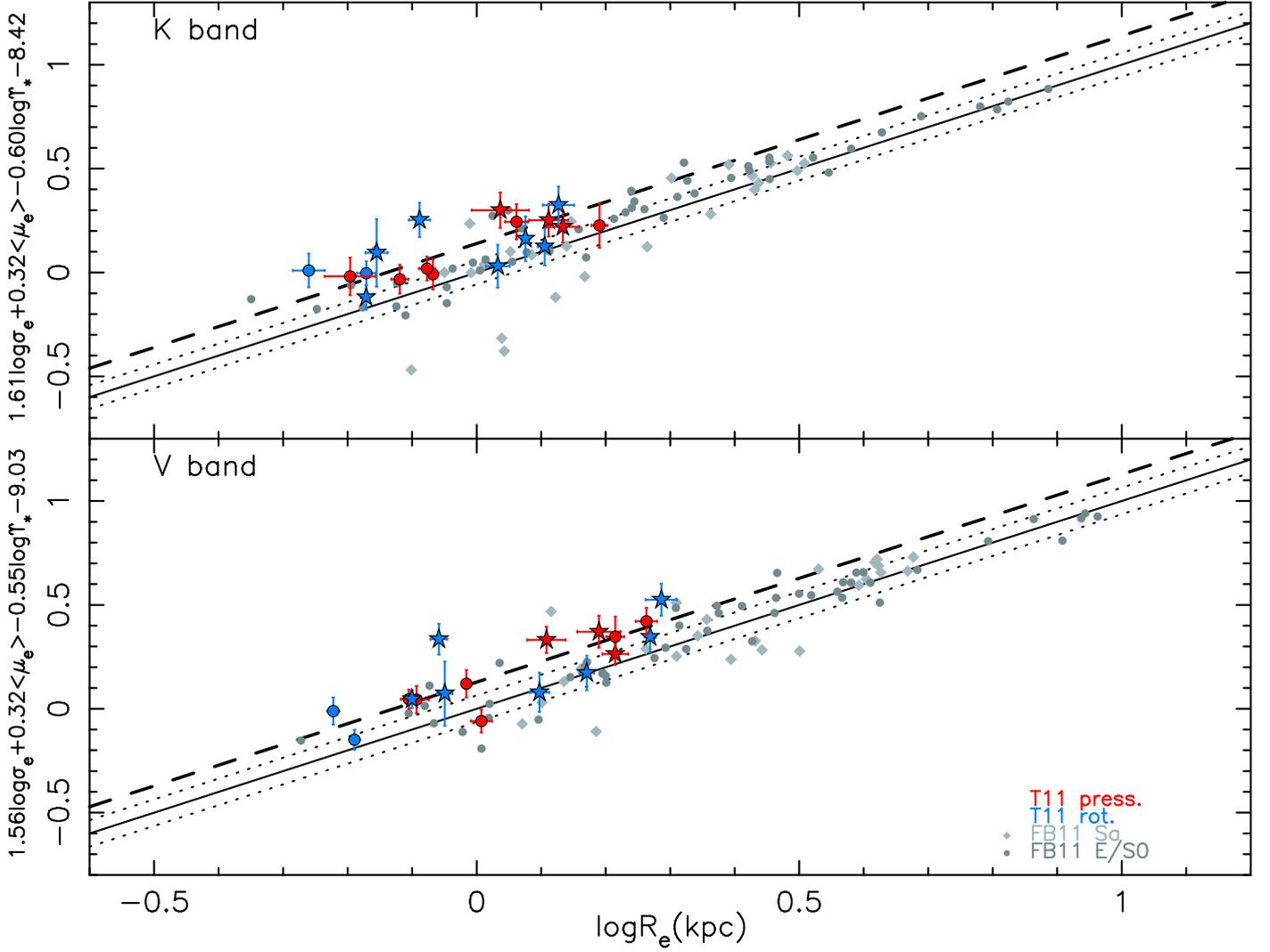}}
\caption{Fundamental Plane in the $K$ and $V$ bands when the stellar mass-to-light ratio ($\Upsilon_*$) is included. Symbols as in Figure \ref{Col-sig}.}
              \label{FP_ML}
\end{figure*}

One can think of two major effects responsible for the offset that dEs show with respect to the plane defined by massive early-type systems: 1) differences in stellar populations, 2) differences in dark matter fractions.

FB11 found that those galaxies with extremely young populations, and signs of ongoing extended star formation, have the largest offset and are displaced below the FP. This is not the case of our sample of dEs: although some of them are as young as $\sim 1$ Gyr, they do not show emission lines, and their offset is above the plane, not below it. Moreover, not only those with the youngest populations are away from the plane of Es, but all of them are. Even so, those that are older are, on average, more separated from the plane than those that are younger. For these reasons, a pure stellar population effect is discarded.

Several studies have tried to explain the thickness of the FP as a mass-to-light ratio effect \citep[see e.g.][FB11]{Zaritsky06,Zaritsky08,Grav10a}. Following the idea of the Fundamental Manifold \citep{Zaritsky06}, FB11 include a new term in the FP, the stellar mass-to-light ratio ($\Upsilon_*$), and they find that those galaxies that had the largest deviation with respect to the plane move closer to it.

To measure $\Upsilon_*$ we generate a grid of 11 models of exponentially declining star formation histories with values of $\tau$, the declining time scale, from 0.1 to 10.0 based on the MILES stellar library \citep{SB06lib,Vazdekis10}  and assuming a Kroupa initial mass function (IMF), as done for the Es. This is based on the fact that the star formation history of dEs has been extended over time \citep[e.g.][]{Boselli08a,Mich08}. For each of the models we get $\Upsilon_*$, in the $V$ and $K$ bands, and the $H_{\beta}$ line strength in the Lick system. We find the best linear fit for the $\Upsilon_* - H_{\beta}$ relation obtained from the models and apply it to the $H_{\beta}$ values of our dEs by \citet{Mich08}.

Figure \ref{FP_ML} shows the FP when $\Upsilon_*$ is included. The grey dashed line, the averaged perpendicular distance of the dEs to the plane, indicates that a systematic offset still remains. Table \ref{dists_rels} shows that, with respect to the classic FP, the offset of rotationally and pressure supported dEs has been inverted, in the sense that the rotationally supported dEs are now those with the largest offset from the plane, but still the pressure supported dEs have the lowest scatter. This offset, $d_{\bot,K}=0.10 \pm 0.07$ and $d_{\bot,V}=0.09 \pm 0.08$, can be used to estimate the dark matter fraction of dEs within their $R_e$, relative to that for Es, by rewriting the FP relation as:

\begin{equation}
\log R_e= \alpha ~  \log \sigma_e + \beta ~ <\mu_e> + \gamma + \delta ~ \log \Big ( \Upsilon_*\frac{M_{\rm dyn}}{M_*}\Big )
\end{equation}

Transforming the measured perpendicular distances along the y-axis we get in this way lower limits on the total-to-stellar-mass fractions of $M_{\rm dyn}/M_*=1.7 \pm 0.5$ in the $K$ band, and $M_{\rm dyn}/M_*=1.7 \pm 0.6$ in the $V$ band. 
As mass ratios ought to be independent of wavelength range, it is indeed good to see that both values are consistent with each other, as well as the independent estimate of $M_{\rm dyn}/M_*=1.6 \pm 1.2$ obtained in Paper I. Whereas here we used the FP offset combined with an estimate of the stellar mass contribution based on exponentially declining star formation and the measured Hbeta line strength, in Paper I we used the virial mass estimate of \citet{Cappellari06} combined with simple stellar population modeling using the age and metallicity as derived by \citet{Mich08}.

\section{Discussion and conclusions}\label{FP_disc}

\begin{figure*}
\centering
\resizebox{0.7\textwidth}{!}{\includegraphics[angle=-90]{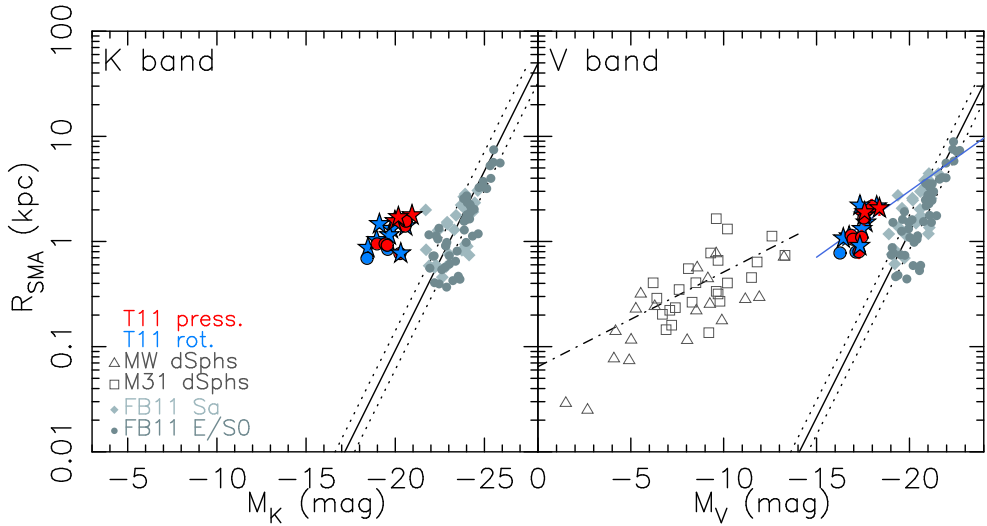}}
\caption{Size-luminosity relation. Symbols as in Figure \ref{Col-sig} including the dSphs from the Milky Way (grey triangles) and M31 (grey squares). The blue line in the $V$ band is the fit by \citet{Shen03} for late-type galaxies. 
The dotted dashed line is the fit for dSphs by \citet{Brasseur11}. All the half light radii have been transformed to semi-major radii, but the fit (solid line) and $1\sigma$ scatter (dotted lines) for the SAURON early-type galaxies from FB11 is based on geometric radii.}
              \label{SL_rels}
\end{figure*}

Scaling relations have been historically used as an observational constraint in the study of galaxy formation and evolution.
The aim of this paper is to extend previous studies, mainly dedicated to massive Es, to dEs, with particular attention to possible systematic differences between pressure and rotationally supported systems.
We have focused this paper on the kinematic scaling relations: the colour-velocity dispersion, the Faber-Jackson, and the Fundamental Plane relation. To investigate where dEs lie with respect to Es, we use the scaling relations of FB11 for the E, S0 and Sa galaxies of the SAURON survey.

In the case of  $(V-K)_e$ versus velocity dispersion, the dEs extend a much larger range in colour than Es, making them to be above the fit for more massive objects. While in the case of Sa galaxies their large dynamic range in $(V-K)_e$ can be explained as the presence of dust, dEs do not show any indication for the presence of dust. Moreover, based on Figure 7 by FB11, this is not a metallicity effect. To understand the behaviour of the optical-infrared colours, more and better data will have to be obtained. 

Regarding the Faber-Jackson relation, we find that dEs, although within the 1$\sigma$ scatter of Es, tend to lie below them, in the same direction that dSphs deviate. This result might disagree with other authors, such as \citet{Matkovic05}, who found that Es and dEs lie on the same Faber-Jackson relation.

In the analysis of the FP we find that dEs are clearly offset with respect to the plane defined by Es. This result is in agreement with \citet{DR05}. They suggest that different star formation histories can explain the offsets from the plane of massive ellipticals. However, plotting the FP in $\kappa$-space, we have realized that while dEs follow the same surface brightness and mass relation as Es, that can even be extended to dSphs, they show an offset in their $M_{\rm dyn}/L$ ratio: they have values that are larger than that of Es, but still lower than the ones of dSphs. This suggests that their location with respect to the plane and Faber-Jackson diagrams is for a large part related to their dynamical mass-to-light ratio \citep[][]{Grav10a}. The offset found with respect to the FP, that is not explained by stellar populations, is directly related to $M_{\rm dyn}/M_*$. Therefore, this offset can be used to estimate the dark matter content of dEs relative to Es. The values found for this ratio are $M_{\rm dyn}/M_* \ge 1.7 \pm 0.5$ in the $K$ band, and $M_{\rm dyn}/M_* \ge 1.7 \pm 0.6$ in the $V$ band. These values, fully consistent with $1.6 \pm 1.2$ measured in Paper I in the $I$ band and with a completely independent method, indicate that the dark matter (DM) fraction of dEs is not as high as the one for dSphs, but it is still measurable: $M_{\rm DM}/M_{\rm dyn} \gtrsim 42\%$ within their half light radius (uncertainties of 17$\%$ in the $K$ band and 20$\%$ in the $V$ band).

From the study of these relations one can see that dEs have properties that are different from Es. This is also seen from the size-luminosity relation in Figure \ref{SL_rels}, where dEs do not follow the same trend as Es \citep[see also][]{Kormendy85,Janz08,Kormendy09,Kormendy12}. In the $V$ band dEs follow the same relation as late-type galaxies and this can, within the uncertainties, be extended towards the faintest dSphs. Although no infrared photometry is available for these dSphs, the different behaviour between Es and dEs is visible in the $K$ band, making this result independent from the passband and the stellar populations of the galaxies involved. \citet{Brasseur11} argue that this common size-luminosity relation from dSphs to late-type galaxies is related to the initial baryonic angular momentum. The fact that we find that dEs follow the same relation as late-type galaxies indicates that they also conserve the angular momentum of their progenitors. This scenario agrees with our finding in Paper I, where rotationally supported dEs have the same shape of the rotation curve as late-type galaxies of similar luminosity. Therefore, the scenario of late-type galaxies being transformed into dEs is further supported. 

The study of the scaling relations strengthens the argument that the properties of dEs are different from those of Es, suggesting that their formation mechanisms are independent. While it is widely accepted that massive early-types are likely formed by mergers \citep[e.g.][]{BoylanKolchin06,Robertson06}, the simulations performed for dwarf galaxies reject this possibility \citep{deLucia06}, and the results found here and in other works \citep[e.g.][]{Kormendy85,Bender92,Kormendy09, etj09b,etj10, Kormendy12} indicate that dEs are the result of transformations of late-type galaxies \citep[but see e.g.][]{Graham97,Graham03,Ferrarese06}. We expect that, if the dEs are in different states of their transformation this would show up in the scaling relations. We can check that by looking at the various subgroups of dEs, and, within the errors, there is no difference in the location of rotationally and pressure supported dEs in these relations, as previously found in the study of the Fundamental Plane in the $\kappa$-space by G03. The only appreciable distinction between the two populations is that pressure supported dEs have the lowest intrinsic scatter.

We have to remark that the sample of pressure and rotationally supported dEs is not complete in any sense but biased towards the brightest dE members of the Virgo cluster. They might thus not be  representative of the whole dE populations of the cluster.
In order to improve the statistical significance of this analysis we are conducting new spectroscopic and photometric observations of dEs in the Virgo cluster to achieve a complete and representative sample of this population of objects \citep[see][]{Janz12}.

\begin{acknowledgements}
We thank the anonymous referee for his/her comments that have helped us to significantly improve this manuscript. ET thanks Juan Carlos Mu\~{n}oz-Mateos, Armando Gil de Paz, Mina Koleva, Sven de Rijcke, Puragra GuhaThakurta and Joshua Simon for very useful discussions.
We thank the MAGPOP EU Marie Curie Training Network for financial support for the collaborating research visits and observations that allowed to make this paper. The observing time used in this work was part of the International Time Program (ITP 2005-2007, ITP4) at El Roque de los Muchachos Observatory (La Palma, Spain). ET thanks the financial support of the Fulbright Program jointly with the Spanish Ministry of Education. ET and JG have received financial support through the Spanish research projects AYA2007-67752-C03-03 and AYA2010-21322-C03-03. JFB acknowledges support from the Ram\'{o}n y Cajal Program as well as grant AYA2010-21322-C03-02 by the Spanish Ministry of Science and Innovation. This paper made use of the following public databases: SDSS \citep{SDSS}, NASA/IPAC Extagalactic Database (NED, operated by the Jet Propulsion Laboratory, California Institute of Technology, under contract with the National Aeronautics and Space Administration), 2MASS \citep[Two Micron All Sky Survey,][which is a joint project of the University of Massachusetts and the Infrared Processing and Analysis Center/California Institute of Technology, funded by the National Aeronautics and Space Administration and the National Science Foundation.]{2MASS}, HyperLEDA \citep{Paturel03}, GOLDMine \citep{GOLDMine}.
\end{acknowledgements}

\bibliographystyle{aa}
\bibliography{references}{}

\begin{appendix}

\section{Errors in the asymptotic magnitudes}\label{app_errors}

The errors of the optical magnitudes, here expressed in Vega system using the SDSS AB to Vega transformation of \citet{Blanton07}, have been computed as described in Appendix A of Paper I. 

The errors of the near-infrared magnitudes have been calculated following the description of \citet{JC09}, based on \citet{GdP05}. Calculating the magnitude as

\begin{equation}
m=-2.5log(F-F_{sky})+ZP
\end{equation}

\noindent where $F$ is the total flux measured in the image, $F_{sky}$ is the flux of the sky, and $ZP$ is the zero point flux calibration. To the first order, the error in the magnitude can be obtained as

\begin{equation}
\Delta m =\sqrt{\Big(\frac{2.5log(e)}{F-F_{sky}}\Big)^2(\Delta F^2+\Delta F_{sky}^2)+\Delta ZP^2}
\end{equation}

\noindent where $\Delta F$, $\Delta F_{sky}$ and $\Delta ZP$ are the errors in the incident flux, the sky flux and the zero point flux calibration respectively.

The uncertainty in the incident flux can be estimated assuming Poissonian statistics 

\begin{equation}
\Delta F=\sqrt{F/g_{eff}}
\end{equation}

\noindent where $g_{eff}$ is the effective gain that converts the incoming flux in counts into electrons. 

The uncertainty in the sky flux has two different contributions, from high and low spatial frequency variations. The former is due to the combination of Poisson noise in the $F_{sky}$ and the pixel-to-pixel flat fielding errors, and the latter is due to large-scale flat fielding errors such as reflections or gradients in the background. To estimate these two contributions we have measured the sky in $\sim$10 square regions of $N_{region}$ pixels each, randomly placed around each galaxy far enough from it to avoid contamination from the galaxy itself. $F_{sky}$ is, therefore, the mean sky value in all the boxes and its error can be expressed as

\begin{equation}\label{Fsky_err}
\Delta F_{sky}= \sqrt{N<\sigma_{sky}>^2+N^2{\rm max}\Big( \sigma_{<sky>}^2-\frac{<\sigma_{sky}>^2}{N_{region}},0 \Big)}
\end{equation}

\noindent where  $<\sigma_{sky}>$  is the mean standard deviation, $\sigma_{<sky>}$ is the standard deviation of the mean sky values among different boxes, and $N$ is the number of pixels used to compute the mean flux within each isophote.
The second term of Equation \ref{Fsky_err}, that accounts for the large-scale background errors, can be neglected if the low-frequency flat-fielding errors are negligible compared with the combined effect of the sky photon noise and the high-frequency flat fielding errors, but this is not usually our case, with the sky background often being not flat.

\section{The Kormendy relation, the colour-magnitude and the colour-size diagrams}\label{app_phot_rels}

These relations are shown in this Appendix for completeness and are similar to those already published in the literature \citep[e.g.][]{Ferrarese06,Lisk08,Boselli08b,Kormendy09,Kormendy12}.

\begin{figure}
\centering
\resizebox{0.45\textwidth}{!}{\includegraphics[angle=-90]{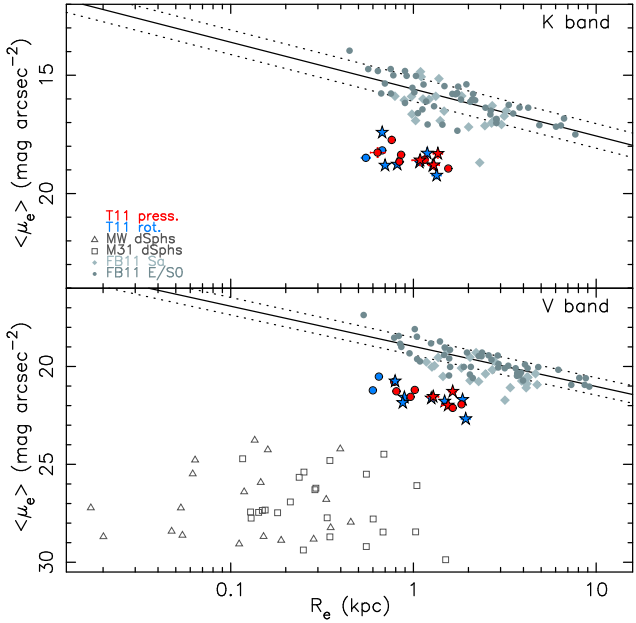}}
\caption{Kormendy relation in $K$ band, upper panel, and $V$ band, lower panel. Symbols as in Figure \ref{Col-sig}. Solid and dotted lines are the fit and $1\sigma$ scatter for the early-types of the SAURON sample of galaxies.}
              \label{Kormendy}
\end{figure}

Figure \ref{Kormendy} presents the Kormendy relation for massive and dwarf early-type galaxies. In both the upper and lower panels, $K$ and $V$ bands respectively, the trends observed are very similar. Whereas the SAURON sample of galaxies follows a correlation where the brightest objects are the smallest ones, dEs present an offset with respect to them, having a lower surface brightness for their size. This is in the direction of dSph galaxies, showing a large range of sizes and lower surface brightnesses.

\begin{figure*}
\centering
\resizebox{0.85\textwidth}{!}{\includegraphics[angle=-90]{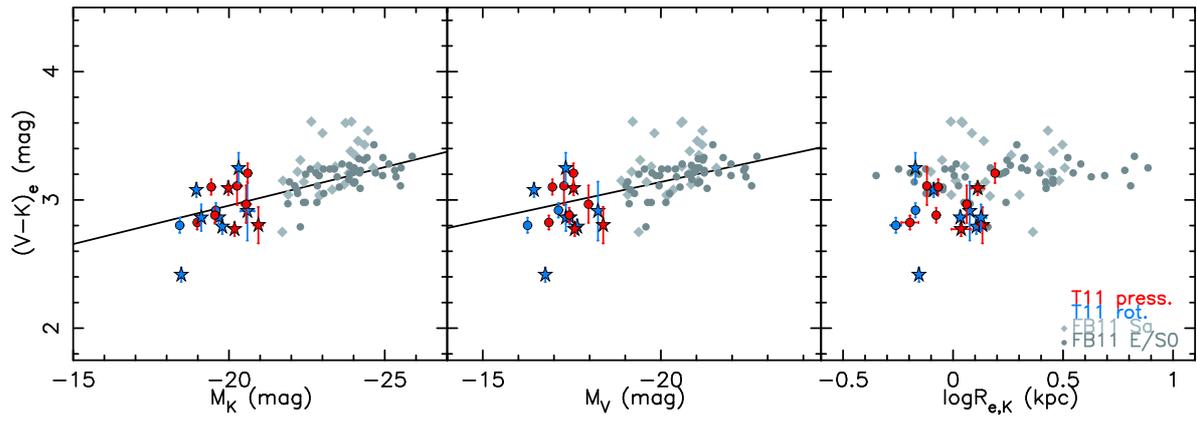}}
\caption{Colour-magnitude and colour-size diagrams. Symbols as in Figure \ref{Col-sig}.}
              \label{CMD}
\end{figure*}

Figure \ref{CMD} shows the colour-magnitude diagram in the infrared (left panel), in the $V$ band (central panel) and the colour-size relation (right panel), the fits plotted are those for massive early-type galaxies (FB11). The dEs show a larger dispersion than the massive early-types, a dispersion that is more similar to the one presented by Sa galaxies. This large dispersion is likely due to the luminosity-metallicity relation of passive galaxies reported by \citet{Grav09}, but due to the large errors in the metallicities for our sample of galaxies \citep{Mich08} we are unable to verify this.

In the case of the colour-size relation, all the samples overlap in the same range of sizes, but the dwarf systems are systematically bluer than the massive galaxies \citep[as previously found in the colour-magnitude relation by e. g.][]{Boselli05,Ferrarese06,Janz09}, indicating younger stellar populations \citep{Mich08}.

We do not see any significant difference between rotationally and pressure supported dEs  in any of these diagrams.

\end{appendix}

\end{document}